\documentclass[epj]{svjour}
\usepackage{graphics,amsmath,enumitem,array,multirow,bm,ulem,enumitem,slashed,xcolor,setspace,booktabs,hyperref,cite,url,tabularx, float}
\graphicspath{ {Figures/} }
\begin{document}
%
\title{Discovery Prospects for the 150 GeV charged scalar at Future $e^+e^-$ Colliders}
\author{
Siddharth P. Maharathy
\inst{1,2,3,}\thanks{\email{siddharth.prasad.maharathy@cern.ch}} \and 
Phodiso Maroeshe
\inst{1,2,}\thanks{\email{phodiso.maroeshe@cern.ch}}
\and 
Paballo Ndhlovu
\inst{1,2,}\thanks{\email{baballo-victor.ndhlovu@cern.ch}}
\and 
Srimoy Bhattacharya
\inst{1,2,}\thanks{\email{srimoy.bhattacharya@cern.ch}}
\and 
Andreas Crivellin
\inst{4,}\thanks{\email{andreas.crivellin@physik.uzh.ch}}
\and 
Mukesh Kumar
\inst{1,}\thanks{\email{mukesh.kumar@cern.ch}}
\and 
Rachid Mazini
\inst{1,}\thanks{\email{rachid.mazini@cern.ch}}
\and 
Bruce Mellado
\inst{1,2,}\thanks{\email{bmellado@mail.cern.ch}}
}                     
%
\institute{School of Physics and Institute for Collider Particle Physics, University of the Witwatersrand, Johannesburg, Wits 2050, South Africa  
\and iThemba LABS, National Research Foundation, PO Box 722, Somerset West 7129, South Africa
\and Indian Institute of Science Education and Research Pune, Dr. Homi Bhabha Road, Pune 411008, India
\and Physik-Institut, Universität Zürich, Winterthurerstrasse 190, CH–8057 Zürich, Switzerland}
%
%
\abstract{
The Real Higgs Triplet model, known as the $\Delta$SM, is a minimal extension of the Standard Model (SM) obtained by adding a hypercharge 0 triplet ($\Delta$). This simple model is motivated by the multi-lepton anomalies and excesses in di-photon, $Z\gamma$, and $WW$ spectra at $\approx152$\,GeV. The model contains, in addition to the SM particle content, a $CP$-even neutral Higgs ($\Delta^0$) and a charged state ($\Delta^\pm$), which are quasi-degenerate in mass. Observing the charged scalar at the LHC and measuring its mass is very challenging, since it dominantly decays to $WZ$, $tb$, and $\tau\nu$. In this article, we consider the discovery prospects of the  charged Higgs with mass 150 GeV at future electron-positron colliders. Taking into account $e^+e^- \to \gamma^*,Z^* \to \Delta^\pm \Delta^\mp$ as the production mechanism and the dominant decay modes, we define three signal regions (SR) to study the 150 GeV charged Higgs properties: SR1: $\ge 3j + 1\ell$, SR2: $\ge 3\ell + \tau_{\text{had}}$, SR3: $\ge 4j + \tau_{\text{had}}$. For $m_{\Delta^\pm}=150\,\text{GeV}$, a $5\sigma$ significance can be achieved in SR1 with an integrated luminosity of less than $1\text{ fb}^{-1}$. SR2 is very clean with leptonic final states having low background and small systematic uncertainties. Furthermore, SR3 is crucial for reconstructing the charged scalar invariant mass, which can be measured with $\mathcal{O}(1)$\,GeV accuracy with an integrated luminosity of $500\text{ fb}^{-1}$.
} 
\date{}
\maketitle
 \section{Introduction}\label{sec:intro}

The Brout-Englert-Higgs (BEH)~\cite{Higgs:1964ia,Englert:1964et,Higgs:1964pj,Guralnik:1964eu} mechanism implemented within the Standard Model (SM)~\cite{GLASHOW1961579, Weinberg:1967tq, Salam:1968rm} has been confirmed by the discovery of a 125\,GeV scalar boson at the Large Hadron Collider (LHC)~\cite{Aad:2012tfa,Chatrchyan:2012ufa}. Since this completed the quest for the last missing piece of the SM, the observation of any new fundamental particle would clearly prove the existence of physics beyond the SM (BSM). In fact, it is clear that the SM cannot be the ultimate theory of Nature since it cannot account for several experimental observations, such as the existence of dark matter and neutrino oscillations, and must therefore be extended. However, the exact field content and structure of a theory that supersedes the SM have yet to be determined.

Given the absence of any symmetry that protects the minimality of the SM Higgs sector with a single $\mathrm{SU}(2)_L$ doublet, extensions of the scalar sector are theoretically well motivated in light of its internal consistency problems, such as vacuum stability, the hierarchy problem, and the flavour puzzle. In fact, the literature discusses a wide range of models that extend the SM scalar sector by incorporating $SU(2)_L$ singlets~\cite{Silveira:1985rk,Pietroni:1992in,McDonald:1993ex}, doublets~\cite{Lee:1973iz,Haber:1984rc,Kim:1986ax,Peccei:1977hh,Turok:1990zg}, and triplets~\cite{Konetschny:1977bn,Cheng:1980qt,Lazarides:1980nt,Schechter:1980gr,Magg:1980ut,Mohapatra:1980yp} etc. These theories are viable even though the discovered 125\,GeV Higgs has properties consistent with the SM predictions, if their contribution to electroweak symmetry breaking is sufficiently small. 

In this context, interestingly, statistically significant indications for new Higgs bosons have been observed, called the multi-lepton anomalies -- i.e.~deviations from the SM predictions in channels with multiple leptons, missing energy and possibly ($b$-)jets~\cite{vonBuddenbrock:2016rmr,vonBuddenbrock:2017gvy,vonBuddenbrock:2019ajh,vonBuddenbrock:2020ter,Hernandez:2019geu,Coloretti:2023wng,Banik:2023vxa,Coloretti:2023yyq,Crivellin:2023zui}. Based on the invariant mass of lepton pairs, the mass of the new scalar was predicted to be $m_S = 150 \pm 5\,\mathrm{GeV}$ (see.~\cite{vonBuddenbrock:2017gvy} for further discussion) and, in fact, Refs.~\cite{Crivellin:2021ubm,Bhattacharya:2023lmu,Bhattacharya:2025rfr} found excesses in the sidebands of SM Higgs searches from ATLAS and CMS experiments~\cite{Sirunyan:2021ybb,ATLAS:2020pvn,Aad:2020ivc,Sirunyan:2020sum,Aad:2021qks,CMS:2018nlv, Sirunyan:2018tbk,ATLAS:2020fcp}. This suggest the presence of a resonance at $m_S \approx 152 \pm 1\,\mathrm{GeV}$, with a global significance of $5.3\sigma$~\cite{Bhattacharya:2025rfr}.  Given the absence of an excess in the $ZZ$ final state, the real Higgs triplet model ($\Delta$SM), involving a $Y = 0$ scalar triplet, provides a well-motivated framework to account for the observed anomalies. 

A detailed analysis of the $\Delta$SM has been carried out in Refs.~\cite{Ashanujjaman:2024pky,Crivellin:2024uhc,Ashanujjaman:2024lnr}, investigating the possible contribution of the Drell-Yan production of the neutral component of the triplet ($\Delta^0$) at the LHC. However, observing the charged component of the triplet at the LHC is difficult due to several challenges. First of all, the VBF production cross section is suppressed by the smallness of the triplet's vacuum expectation value, and the main production channel is Drell-Yan with a cross section below the picobarn level. Furthermore, for low masses ($m_{\Delta^\pm}<120$\,GeV), the $\Delta^\pm$ dominantly decays to $\tau^\pm \nu$. This resembles the signatures of stau-like supersymmetric particles decaying to a $\tau$ and a (massless) neutralino, making them hard to isolate due to the missing energy in the final state. Thus, using the CMS stau search~~\cite{CMS:2022syk} only a lower bound of $m_{\Delta^\pm}> 110$\,GeV can be set~\cite{Ashanujjaman:2024lnr}. Additionally, for heavier masses, the dominant decay is the $WZ$ channel (with one of the $W$ or $Z$ being possibly off-shell). Here, we have either low-energetic jets or missing energy, and current data only allows excluding the mass region between $\approx 200\,$GeV -- 220\,GeV~\cite{Butterworth:2023rnw}.\footnote{Note that the small mass splitting between $\Delta^\pm$ and $\Delta^0$ enforced by electroweak precision data and perturbativity, disfavours the cascade decays.} 

Given the limitations in observing $\Delta^\pm$ at the LHC, we consider in this article the possibility of discovering $\Delta^\pm$ at future $e^+e^-$ colliders, like the Circular Electron-Positron Collider (CEPC)~\cite{CEPCStudyGroup:2018ghi,An:2018dwb}, the Compact Linear Collider (CLIC)~\cite{CLICdp:2018cto}, the Future Circular Collider (FCC-ee)~\cite{FCC:2018evy,FCC:2018byv} and the International Linear Collider
(ILC)~\cite{ILC:2013jhg,Adolphsen:2013jya}. These machines provide a clean experimental environment with significantly lower backgrounds compared to hadron colliders, enhancing the sensitivity to electroweak-scale bosons with soft particles in the final states.

The remainder of this article is organized as follows. In \ref{sec:model}, we present the framework of the real Higgs triplet model ($\Delta$SM), introduce the scalar potential, and discuss the resulting mass spectrum and parameter space. In \ref{sec:decay}, we study the production mechanisms and dominant decay modes of the charged scalar $\Delta^\pm$ at future $e^+e^-$ colliders. In \ref{sec:production}, we perform a detailed collider analysis of the charged scalar signals, defining three representative signal regions. Finally, in \ref{sec:conclusion}, we summarise our findings and highlight the discovery potential of future lepton colliders in probing the charged scalar sector of the $\Delta$SM.

\section{Model Description}\label{sec:model}

We study the Standard Model extended by an $SU(2)_L$ triplet scalar ($\Delta$) with hypercharge $Y=0$, known as the $\Delta$SM model~\cite{Ross:1975fq,Gunion:1989ci,Chankowski:2006hs,Blank:1997qa,Forshaw:2003kh,Chen:2006pb,Chivukula:2007koj,Bandyopadhyay:2020otm}. The scalar fields can be decomposed as:
\begin{align}
\Phi &= \begin{pmatrix} h_\Phi^+ \\ \frac{1}{\sqrt{2}} (v_\Phi + h_\Phi^0 + iG^0) \end{pmatrix}, \label{eq:doublet} \\[1em]
\Delta &= \frac{1}{2} \begin{pmatrix}  v_\Delta + h_\Delta^0 & \sqrt{2}h_\Delta^+ \\ \sqrt{2}h_\Delta^- & -(v_\Delta + h_\Delta^0) \end{pmatrix},\label{eq:triplet}
\end{align}
where $h_{\Phi,\Delta}^0$ are real scalar fields, $h_{\Phi,\Delta}^- = (h_{\Phi,\Delta}^+)^*$, and $v_\Phi \approx 246$\,GeV and $v_\Delta$ are the respective vacuum expectation values (VEVs). 

The scalar potential is given by
\begin{align}
\label{eq:pot}
V = & -\mu_\Phi^2 \Phi^\dag \Phi + \frac{\lambda_\Phi}{4} \left(\Phi^\dag \Phi\right)^2 - \mu_\Delta^2 {\rm Tr}\left(\Delta^\dag \Delta\right) 
\\ \nonumber
& + \frac{\lambda_\Delta}{4} \left[ {\rm Tr}\left(\Delta^\dag \Delta\right) \right]^2  + A \Phi^\dag \Delta \Phi + \lambda_{\Phi \Delta} \Phi^\dag \Phi {\rm Tr}\left(\Delta^\dag \Delta\right),
\end{align}
and minimization of it requires
\begin{align*}
\mu_\Phi^2 &= -\frac{A v_\Delta}{2}+\frac{1}{4} v_\Phi^2 \lambda _{\Phi }+\frac{1}{2} \lambda _{\Phi \Delta } v_\Delta^2,
\\
\mu_{\Delta}^2 &=  -\frac{A v_\Phi^2}{4 v_\Delta}+\frac{1}{2} v_\Phi^2 \lambda _{\Phi \Delta }+\frac{1}{4} \lambda _{\Delta } v_\Delta^2.
\end{align*}
Substituting these expressions for $\mu_\Phi$ and $\mu_{\Delta}$ into the Lagrangian simplifies the $CP$-even and charged scalar mass matrices as
\begin{align*}
M^2_0 &= \begin{pmatrix}
\frac{\lambda_\Phi v_\Phi^2}{2} & \left( \lambda _{\Phi \Delta } v_\Delta-\frac{A}{2} \right) v_\Phi \\
\left( \lambda _{\Phi \Delta } v_\Delta-\frac{A}{2} \right) v_\Phi & \frac{\lambda_\Delta v_\Delta^2}{2} + \frac{Av_\Phi^2}{4v_\Delta} \\
\end{pmatrix},\\
M^2_\pm &= \begin{pmatrix}
A v_\Delta & \frac{A v_\Phi}{2} \\
\frac{A v_\Phi}{2} & \frac{A v_\Phi^2}{4 v_\Delta}
\end{pmatrix},
\end{align*}
in the interaction basis $(h^{0,\pm}_\phi,h^{0,\pm}_\Delta)$. The mass matrices are diagonalised by rotating to the physical basis via
\begin{align*}
\begin{pmatrix} h \\ \Delta^0 \end{pmatrix} &= \begin{pmatrix} \cos\alpha & \sin\alpha \\ -\sin\alpha & \cos\alpha \end{pmatrix} \begin{pmatrix} h_\Phi^0 \\ h_\Delta^0 \end{pmatrix},
\\
\begin{pmatrix} G^\pm \\ \Delta^\pm \end{pmatrix} &= \begin{pmatrix} \cos\beta & \sin\beta \\ -\sin\beta & \cos\beta \end{pmatrix} \begin{pmatrix} h_\Phi^\pm \\ h_\Delta^\pm \end{pmatrix},
\label{Eq:gague_to_physical}
\end{align*}
and the eigenvalues (physical masses) are 
\begin{align}
&m_{h}^2 = \frac{\lambda_\Phi v_\Phi^2}{2} + \tan\alpha \left(\lambda_{\Phi\Delta} v_\Delta -\frac{A}{2}\right) v_\Phi,
\\
&m_{\Delta^0}^2 = \frac{\lambda_\Delta v_\Delta^2}{2} + \frac{Av_\Phi^2}{4v_\Delta} - \tan\alpha \left(\lambda_{\Phi\Delta} v_\Delta -\frac{A}{2}\right) v_\Phi, \label{eq:mH0}
\\
&m_{\Delta^\pm}^2 = A \frac{v_\Phi^2+4v_\Delta^2}{4v_\Delta}, \label{eq:mHp}
\end{align}
while the mixing angles read
\begin{align}
& \tan 2\alpha = \dfrac{4v_\Phi v_\Delta \left(2\lambda_{\Phi\Delta} v_\Delta -A\right)}{2\lambda_\Phi v_\Phi^2 v_\Delta -2\lambda_\Delta v_\Delta^3 -A v_\Phi^2},
\\
& \tan 2\beta = -\dfrac{4v_\Phi v_\Delta}{v_\Phi^2-4v_\Delta^2}.
\end{align}
We identify $h$ with the SM-like 125\,GeV Higgs observed at the LHC. $G^\pm$ and $G^0$ are the charged and neutral Goldstone bosons, respectively. 

Now, we can write down all the Lagrangian parameters in terms of physical masses, the triplet VEVs, and the mixing angles as
\begin{align*}
& \lambda_\Phi = \frac{2}{v_\Phi^2} \left[ \cos^2\alpha ~m_{h}^2 + \sin^2\alpha ~m_{\Delta^0}^2 \right],
\\
&\lambda_\Delta = \frac{2}{v_\Delta^2} \left[ \sin^2\alpha ~m_{h}^2 + \cos^2\alpha ~m_{\Delta^0}^2 - \frac{v_\Phi^2}{v_\Phi^2+4v_\Delta^2} m_{\Delta^\pm}^2 \right],
\\
&\lambda_{\Phi\Delta} = \frac{1}{2v_\Phi v_\Delta} \left[ \sin 2\alpha \left(m_{h}^2-m_{\Delta^0}^2\right) + \frac{4v_\Phi v_\Delta}{v_\Phi^2+4v_\Delta^2} m_{\Delta^\pm}^2 \right],
\\
&A = \frac{4v_\Delta}{v_\Phi^2+4v_\Delta^2} m_{\Delta^\pm}^2.
\end{align*}
Therefore, we have four free parameters (in addition to the SM ones): $m_{\Delta^0}, m_{\Delta^\pm}, \alpha$ and $v_\Delta$.

The model also impacts electroweak observables via its custodial symmetry breaking VEV, which only contributes to the $W$ mass:
\begin{equation}
m_W \approx m_W^{\mathrm{SM}} \left( 1 + \frac{2 v_\Delta^2}{v_\Phi^2} \right).
\end{equation}
The global fit to electroweak precision data~\cite{ParticleDataGroup:2022pth, Cheng:2022hbo,FileviezPerez:2008bj,Kanemura:2012rs} shows that the world average, including the CDF II result~\cite{CDF:2022hxs}, favours $v_\Delta = 3.4 \pm 1.0$\,GeV, while excluding CDF II leads to a slightly lower central value, $v_\Delta = 2.3 \pm 1.7$\,GeV. Furthermore, for low masses, bounds from top quark decays $t\to b H^\pm$ lead to even more stringent bounds on the VEV~\cite{Ashanujjaman:2025una}. Since from Eq.~\eqref{eq:mH0} and Eq.~\eqref{eq:mHp} we have
\begin{align}
m_{\Delta^\pm}^2 - m_{\Delta^0}^2 = Av_\Delta - \frac{\lambda_\Delta v_\Delta^2}{2} + \tan\alpha \left(\lambda_{\Phi\Delta} v_\Delta -\frac{A}{2}\right) v_\Phi.
\label{eq:mass_splitting}
\end{align}
The neutral and charged scalars, $\Delta^0$ and $\Delta^\pm$, are degenerate in mass in the limit $\alpha \to 0$ and $v_\Delta \to 0$, and using the requirements of vacuum stability and perturbative unitarity, the mass splitting is restricted to be at most of the same order.

\section{Production and Decay of $\Delta^\pm$ at $e^+ e^-$ Colliders}\label{sec:decay}

\begin{figure}[htb!]
\centering
\resizebox{0.35\textwidth}{!}{%
  \includegraphics{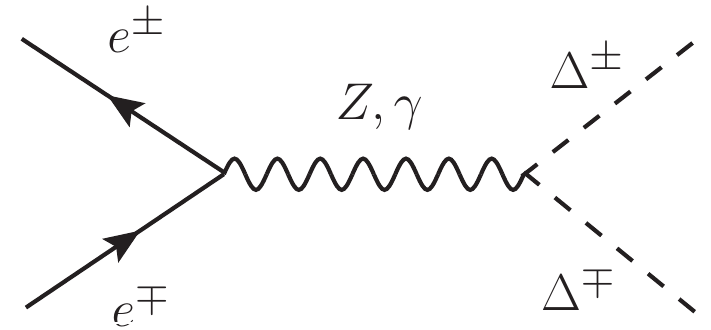}
}
\caption{Feynman diagram showing the Drell-Yan production of the charged scalar ($\Delta^\pm$) at $e^+ e^-$ collider. }
\label{fig:Feynmann_prod}
\end{figure}

\begin{figure}[htb!]
\centering
\resizebox{0.45\textwidth}{!}{%
  \includegraphics{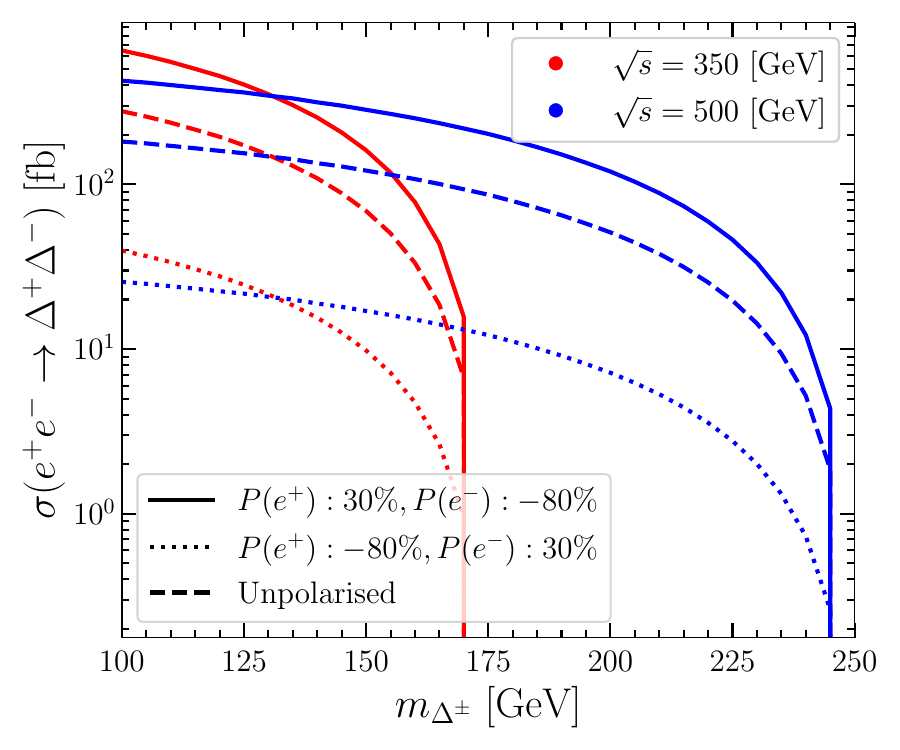}
}
\caption{Cross section for the Drell-Yan production of the charged scalar ($\Delta^\pm$) at $e^+ e^- $ colliders for different center of masses.}
\label{fig:pair production}
\end{figure}

Future $e^+e^-$ colliders, with their clean experimental environment, offer a promising avenue for searching for $\Delta^\pm$ through its electroweak pair production, i.e.,~via Drell-Yan production, Fig.~\ref{fig:Feynmann_prod}. The corresponding cross section is shown as a function of the mass of the charged scalars for two centre-of-mass energies ($\sqrt{s}=350\, \text{GeV}$ and $500\,\text{GeV}$) in Fig.~\ref{fig:pair production} for different polarisations ($P(e^\pm)$) of the electron and positron beams. In our analysis, we have considered the polarization: $P(e^+) = 30\%, P(e^-) = -80\%$.

The experimental signatures are then defined via the subsequent decays of $\Delta^\pm$, where the dominant ones are given by~\cite{Rizzo:1980gz,Keung:1984hn,Djouadi:2005gi,Djouadi:2005gj, Ashanujjaman:2024lnr}:

\begin{align*}
&\Gamma(\Delta^\pm \to ff^\prime) = \frac{N_cm_{\Delta^\pm}^3\sin^2\beta}{8\pi v_\Phi^2} \beta_{ff'}\left(\frac{m_f^2}{m_{\Delta^\pm}^2},\frac{m_{f'}^2}{m_{\Delta^\pm}^2}\right),
\\
&\Gamma(\Delta^\pm \to t^*\bar{b}/\bar{t^*}b \to W^\pm b\bar{b}) = \frac{3m_t^4 m_{\Delta^\pm} \sin^2\beta}{128\pi^3 v_\Phi^4} \times \\
&\hspace{4.6cm}\beta_t\left(\frac{m_t^2}{m_{\Delta^\pm}^2},\frac{m_W^2}{m_{\Delta^\pm}^2}\right),
\\
&\Gamma(\Delta^\pm \to W^\pm Z^*) = \frac{9g^2\lambda_{\Delta^\pm W^\mp Z}^2}{128\pi^3\cos^2\theta_w m_{\Delta^\pm}} (\frac{7}{12}-\frac{10}{9}\sin^2\theta_W   \\ &\hspace{2.8cm} + \frac{40}{27}\sin^4\theta_W) H\left(\frac{m_W^2}{m_{\Delta^\pm}^2},\frac{m_Z^2}{m_{\Delta^\pm}^2}\right),
\\
&\Gamma(\Delta^\pm \to W^{\pm*} Z) = \frac{9g^2\lambda_{\Delta^\pm W^\mp Z}^2}{256\pi^3 m_{\Delta^\pm}} H\left(\frac{m_Z^2}{m_{\Delta^\pm}^2},\frac{m_W^2}{m_{\Delta^\pm}^2}\right),
\\
&\Gamma(\Delta^\pm \to h W^{\pm *}) = \frac{9g^2m_{\Delta^\pm}}{128\pi^3} \lambda_{\Delta^\pm hW^\mp}^2 G\left(\frac{m_{h}^2}{m_{\Delta^\pm}^2},\frac{m_W^2}{m_{\Delta^\pm}^2}\right),
\end{align*}
where
\begin{align*}
&\lambda_{\Delta^\pm W^\mp Z} = -\frac{g^2}{2\cos\theta_w}\left(2v_\Delta\cos^2\theta_w\cos\beta-v_\Phi\sin^2\theta_w\sin\beta\right),
\\
&\lambda_{\Delta^\pm hW^\mp} = -\frac{g}{2}\left(2\sin\alpha \cos\beta + \cos\alpha \sin\beta \right).
\end{align*}
$g$ is the $SU(2)_L$ gauge coupling and $\theta_w$ is the Weinberg angle. The widths above involve several kinematic functions, which encapsulate the effects of two- and three-body phase space as well as off-shell kinematics. For completeness, the analytical forms of these functions~\cite{Rizzo:1980gz,Keung:1984hn,Djouadi:2005gi,Djouadi:2005gj,Ashanujjaman:2024lnr} are collected in Appendix~\ref{app:LoopFunc}.

Note that since the tree-level induced decays of $\Delta^\pm$ only depend on $m_{\Delta^\pm}$ and $v_\Delta$ (except for the $h W^\pm$ mode which also depends on neutral mixing $\alpha$) the dependence on $v_\Delta$ drops out to a good approximation, as long as $v_\Delta$ is $\mathcal{O}(1)$\,GeV. Furthermore, the width of cascade decay mode $\Delta^\pm \to W^\pm \Delta^0$ depends in addition on the mass splitting $m_{\Delta^\pm} - m_{\Delta^0}$ and is thus suppressed. Finally, all loop-induced decays are small due to the loop factor and can be neglected.

\begin{figure}[t!]
\centering
\resizebox{0.45\textwidth}{!}{%
  \includegraphics{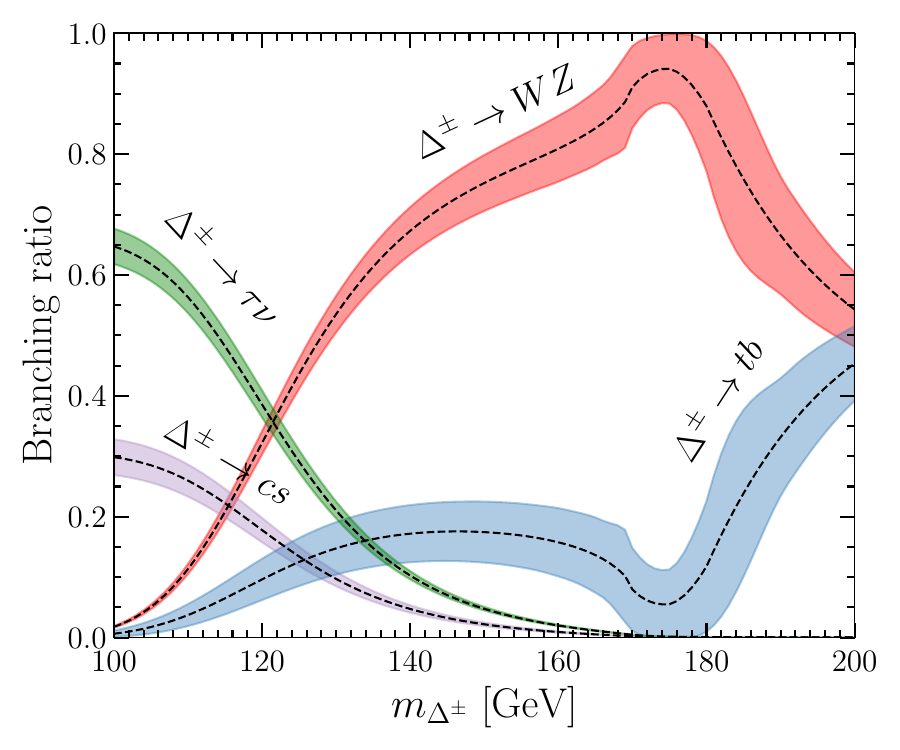}
}
\caption{Dominant branching ratios of $\Delta^\pm$, including the uncertainties estimated from Ref.~\cite{LHCHiggsCrossSectionWorkingGroup:2013rie}, as a function of its mass. We have assumed $\alpha = 0$. }
\label{fig:BrHpm}
\end{figure}

In Fig.~\ref{fig:BrHpm}, we show the dominant branching ratios of $\Delta^\pm$ as a function of $m_{\Delta^\pm}$. The uncertainties are estimated by propagating the errors on the $\tau \tau$, $c\bar{c}$, $t\bar{t}$, and $ZZ^*$ decays of a hypothetical SM-like Higgs with mass $m_{\Delta^\pm}$, as reported in the CERN Yellow Report~\cite{LHCHiggsCrossSectionWorkingGroup:2013rie}. For the respective figure, we have assumed $\alpha = 0$. One can see that the $\tau\nu$ channel dominates at low masses, while the $WZ$ mode becomes the leading channel at intermediate and higher masses, but the $tb$ channel increases at higher masses. 

\section{Discovery Potential at Future $e^+e^-$ Colliders}\label{sec:production}

For our study of the signatures of the charged component of the triplet at future $e^+e^-$ colliders, we consider $m_{\Delta^\pm} = 150 $\,GeV as a benchmark point. This is motivated by the di-photon excess at 152\,GeV and the fact that the charged and neutral components have to be very close in mass. However, note that the general features of this analysis will also hold for other masses at the electroweak scale. We simulated the pair production $e^+ e^- \to \Delta^\pm \Delta^\mp$ at $\sqrt{s} = 350$\,GeV. Motivated by the dominant decay modes of $\Delta^\pm$, we define the following signal regions:
\begin{itemize}
    \item SR1: $\geq 3j + 1\ell$
    \item SR2: $\geq 3\ell + \tau_{\text{had}}$
    \item SR3: $\geq 4j + \tau_{\text{had}}$
\end{itemize}

SR1 is defined such that the signal yield will be large, considering that $\Delta^\pm$ is pair-produced and decays dominantly to $W^\pm Z$. Similarly, the presence of a hadronic final state originating from $W^\pm Z \to 4j$, while targeting $\tau \nu$ decays of the second $\Delta^\pm$, makes it possible to fully reconstruct the invariant mass of the charged scalar $\Delta^\pm$ within SR3. Consequently, SR3 offers a unique avenue not only to enhance the discovery prospects but also to directly measure the mass of the new scalar, which is crucial for establishing the existence of the underlying resonance. However, in SR2, the choice of leptonic final state does not allow mass reconstruction and has lower sensitivity as compared to the earlier SRs. But this SR is very clean with leptonic final states, low background and small systematic uncertainties. 

\begin{table}[t]
\centering
\resizebox{\columnwidth}{!}{%
\begin{tabular}{c c p{6cm}}
\toprule
\multicolumn{2}{c}{\textbf{Signal Region}} & \textbf{$\ell$ and $j$ selection criteria} \\
\midrule
SR1: & $3j+1\ell$ & 
$n_\text{jet} \geq 3$, $n_{\ell=e,\mu} \geq 1$, $p_\text{T}(j) > 20$ GeV, \\
& & $p_\text{T}(\ell) > 10$ GeV, $\theta_{j,\ell} > 9.38^\circ$ \\[0.2cm]

SR2: & $3\ell+\tau_{\text{had}}$ & 
$n_{\ell=e,\mu} \geq 3$, $n_{\tau_{\text{had}}} \geq 1$, $p_\text{T}(\ell) > 10$ GeV, \\
& & $p_\text{T}(\tau_{\text{had}}) > 20$ GeV, $\theta_{\ell,\tau_{\text{had}}} > 9.38^\circ$ \\[0.2cm]

SR3: & $4j+\tau_{\text{had}}$ & 
$n_\text{jet} \geq 4$, $n_{\tau_{\text{had}}} \geq 1$, $p_\text{T}(j) > 10$ GeV, \\
& &$p_\text{T}(\tau_{\text{had}}) > 10$ GeV, $\theta_{j,\tau_{\text{had}}} > 9.38^\circ$ \\
\bottomrule
\end{tabular}
}
\caption{Leptons and jets selection criteria considered for each SR analysed in this work.  $n_{\ell=e,\mu}$, $n_\text{jet}$, $n_{\tau_{\text{had}}}$ stands for the number of leptons, jets and $\tau$-tagged jets, $p_\text{T}$ and $\theta$ denote the transverse momentum and polar angle of the leptons and jets, respectively.}
\label{tab:SRs}
\end{table}

Following event generation with \url{MadGraph5_aMC_v3.5.3}~\cite{Alwall:2011uj, Alwall:2014hca}, parton showering and hadronisation are performed using \url{Pythia 8.3}~\cite{Sjostrand:2014zea} and detector effects are simulated with \url{Delphes 3.5.0}~\cite{deFavereau:2013fsa} (assuming the ILC detector \cite{ILC:2013jhg}). 
The reconstruction of jets, electrons, muons, and missing transverse energy is carried out 
using the Delphes ILD card for $\sqrt{s} = 350$ GeV.  We use the anti-$k_T$ algorithm~\cite{Cacciari:2008gp} implemented in \url{FastJet 3.3.4}~\cite{Cacciari:2011ma} 
for jet clustering. The detailed definition of the SRs along with the preselection requirements are summarized in Table~\ref{tab:SRs}.

As seen from Fig.~\ref{fig:BrHpm}, the dominant decay mode of $\Delta^\pm$ at $m_{\Delta^\pm} = 150$\,GeV is the $W^\pm Z$. For our final states for the three signal regions, the dominant backgrounds originate from di-boson ($VV$), and tri-boson ($VVV$) production, where $V = W^\pm, Z, h$. The presence of a lepton and at least $3j$ in the final state reduces the multi-jet and di-boson background significantly, and hence we do not consider them. 

To improve signal-to-background discrimination, we employ a deep neural network (DNN) for the first two SRs, while for SR3, we use a cut-based analysis. A DNN is a sequential, structured neural network. It has a linear stack of layers, starting with an input layer that reflects the dimension of the input features (observables) used to build the model. We show the input features and their correlations in the appendix. The model architecture of the DNN includes many hidden layers; the specific architecture used will be discussed in later sections, which use the Rectified Linear Unit (ReLU) activation function~\cite{DBLP:journals/corr/abs-1803-08375}:
\begin{align}
    f(x) \equiv max(0,x) = \begin{cases} 
      0 & \text{if } x < 0 \\
      x & \text{if } x \geq 0 
   \end{cases},
\end{align}
where $x$ is the input feature. The model ends with a single-neuron output layer and a sigmoid activation function~\cite{NARAYAN199769}:
\begin{align}
    \sigma_f(x) = \frac{1}{1+e^{-x}},\quad {\rm with}\quad 0 < \sigma_f(x) <1.
\end{align}
Here, $x$ is the input variable for the final output layer, which is the result of the weighted sum of the outputs of the previous layers. To compile the model, we have used the Adam optimiser and the loss function, the binary cross-entropy~\cite{bishop2006pattern}, is defined as:
\begin{align}
-f(y_{\rm true},y_{\rm pred}) =\; & y_{\rm true}\, \log(y_{\rm pred}) \nonumber \\
& + (1 - y_{\rm true})\, \log(1 - y_{\rm pred}),
\end{align}
where $y_\text{true}$ is the actual class label, which is 1 for signal and 0 for background, and $y_\text{pred}$ is the probability of the event to be assigned as signal or background by the DNN architecture.

\subsection{SR1: $\ge 3j + 1\ell$}
\begin{figure}[t!]
\centering
\resizebox{0.4\textwidth}{!}{%
  \includegraphics{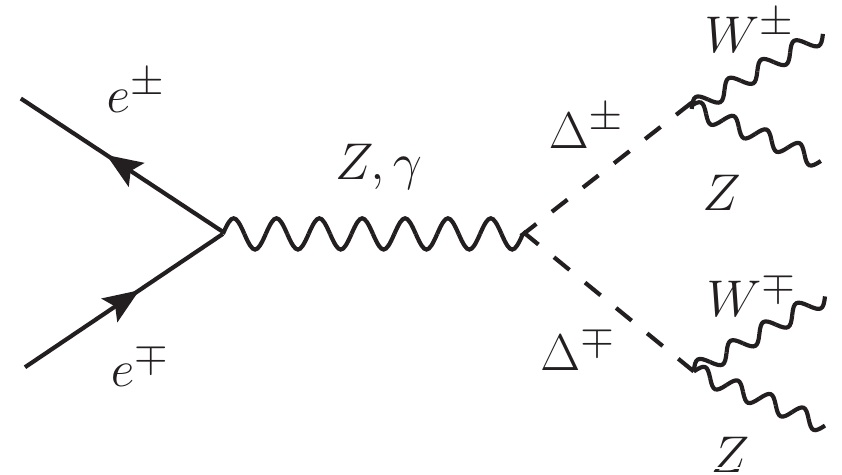}
}

\caption{Feynman diagram for the signal process which contributes to SR1. The produced $W^\pm, Z$ then decay to both jets and leptons.}
\label{fig:Feynmann_Sr1}
\end{figure}

\begin{table}[htb!]
\centering
\resizebox{\columnwidth}{!}{%
\begin{tabular}{p{0.34\columnwidth} p{0.65\columnwidth}}
\toprule
\textbf{Variables} & \textbf{Definition} \\
\midrule
$E(\ell_1)$  & Energy of the leading lepton $(\ell_1)$ \\
$H_\text{T}$  & Scalar sum of the hadronic $p_\text{T}$ \\
$E_{\text{eff}}$  & Effective energy ($= E(\text{jets}) + E(\text{leptons}) + p^{\text{miss}}_\text{T}$) \\
$p_\text{T}(\ell_1)/L_\text{T}$  & Ratio of the leading lepton $p_\text{T}$ over the scalar sum of all leptonic $p_\text{T}$ \\
$\Delta\phi(j_1,\ell_1)$  & Angular separation between the leading jet $(j_1)$ and the leading lepton $(\ell_1)$ \\
$E(j_2)$ & Energy of the sub-leading jet $(j_2)$ \\
$p_\text{T}(j_1)$ & $p_\text{T}$ of the leading jet $(j_1)$ \\
$p_\text{T}(j_1)/H_\text{T}$ & Ratio of the leading jet $p_\text{T}$ and the scalar sum of hadronic $p_\text{T}$ \\
$E(j_3)$ & Energy of the third leading jet $(j_3)$ \\
$E(j_1)$ & Energy of the leading jet $(j_1)$ \\
$M_{\text{eff}}$ & Effective mass ($= H_\text{T} + p^{\text{miss}}_\text{T}$) \\
$\Delta\phi(j_{123},\ell_1)$ & Angular separation between the 3-jet system $(j_{123})$ and the leading lepton $(\ell_1)$ \\
$\Delta\phi(j_2,\ell_1)$ & Angular separation between the sub-leading jet $(j_2)$ and the leading lepton $(\ell_1)$ \\
$m_{j_2, j_3}$ & Invariant mass of the $2^{\rm nd}$ and $2^{\rm rd}$ leading jets $(j_2, j_3)$ \\
$m(j_1,j_2,j_3)$ & Invariant mass of the leading 3-jet system $(j_{123})$ \\
\bottomrule
\end{tabular}%
}
\caption{Input variables used in the DNN training for signal region SR1.}
\label{tab:SR1_variables}
\end{table}

\begin{figure*}[t]
\centering
\resizebox{0.45\textwidth}{!}{%
  \includegraphics{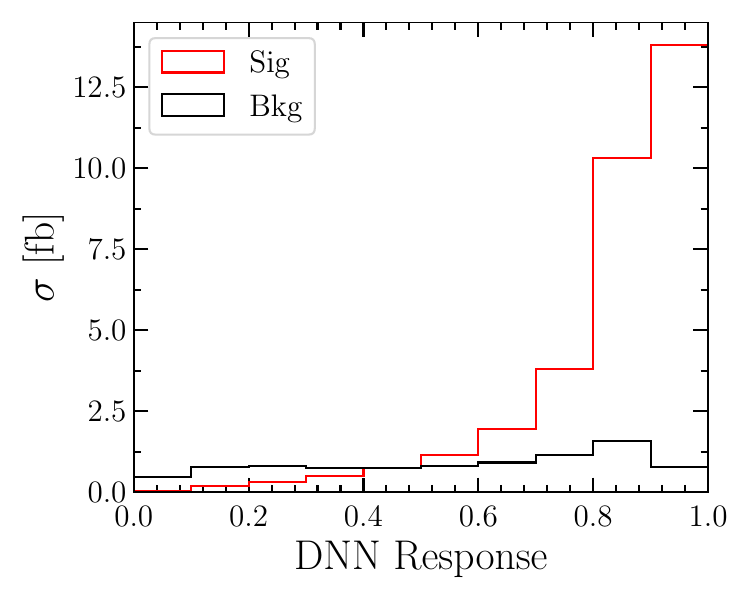}
}
\resizebox{0.45\textwidth}{!}{%
  \includegraphics{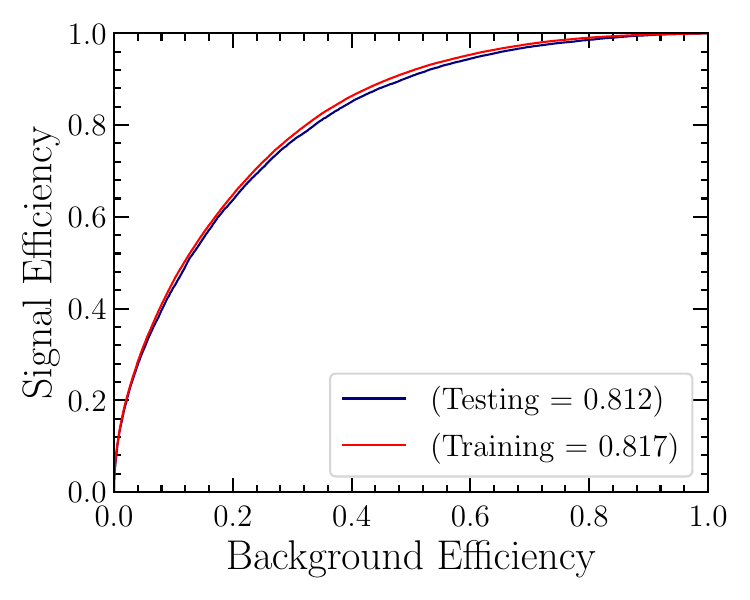}
}
\caption{Left: DNN response for signal and background for SR1. Right: ROC-curve for both training and testing samples for SR1.}
\label{fig:SR1_DNNoutput}
\end{figure*}

In Table~\ref{tab:SR1_variables}, we give the input feature used for DNN for this signal region. We have considered a DNN with 6 hidden layers with 128, 64, 48, 32, 16, and 8 neurons, respectively. Firstly, the data set is divided into a training (80\%) and a testing dataset (20\%). The loss function is computed using the binary cross-entropy function, and the Adam optimiser is used to update weights by backpropagation. After the training, the model is then tested with the rest of the data. The resulting DNN response and Receiver Operating Characteristic (ROC) curves are shown in Fig.~\ref{fig:SR1_DNNoutput}. The Area Under Curve (AUC) scores representing the DNN’s ability to separate
signal and background events and it is found to be 81.2\% and 81.7\% for testing and training samples, respectively. This shows that the DNN model is able to distinguish the signal from the background without overfitting. From the DNN response, Fig.~\ref{fig:SR1_DNNoutput}, we consider a DNN score of 0.7 to be the best score to distinguish the signal from the background. We present the fiducial cross sections of both the signal and background after the above DNN score in Table~\ref{tab:SR1_cutflow}. 

\begin{table}[htbp!]
\centering
\begin{tabular}{lcc}
\toprule
Process &  $\ge 3j + 1\ell$  & DNN Response $>$ 0.7 \\
\midrule
Signal & 32.82 fb & 28.45 fb\\
$VVV$ & 8.76 fb& 3.67 fb\\
\bottomrule
\end{tabular}
\caption{Fiducial cross sections  of both the signal and background after the preselction and DNN score for SR1, where $V = W^\pm, Z \text{ and } h$.}
\label{tab:SR1_cutflow}
\end{table}

Following Refs.~\cite{Cowan:2010js,1983ApJ...272..317L,Cousins:2007yta}, we calculate the expected discovery significance:

\begin{align}
Z_{\rm dis} =& \Bigg[ 2\Bigg( (s+b) \ln \left[ 
   \frac{(s+b)(b+\delta_b^2)}{b^2+(s+b)\delta_b^2} \right] \nonumber \\
& -\frac{b^2}{\delta_b^2} \ln\left [1+ \frac{\delta_b^2 s}{b(b+\delta_b^2)} \right]
   \Bigg) \Bigg]^{1/2},
\label{Eq:significance_formula}
\end{align}
where $s$ and $b$ are the numbers of signal and background events, respectively. $\delta_b$ is the error on the background, which contains uncertainties from object reconstruction, etc. Without delving into the details of estimating $\delta_b$, we conservatively assume it to be 20\%. 
The number of signal ($s$) and background ($b$) events surviving the cuts at a given luminosity $\mathcal{L}$ is given by $s, b = \sigma_{\text{S,B}} \times \mathcal{L}$, where $\sigma_\text{S,B}$ denotes the (background) fiducial cross section. With the values in Table~\ref{tab:SR1_cutflow}, we find that a $5\sigma$ discovery can be achieved with less than $1\text{ fb}^{-1}$ integrated luminosity.

\subsection{\texorpdfstring{SR2: $\ge 3\ell + \tau_{\text{had}}$}{SR2: ≥3l + tau_had}}

\begin{figure}[t!]
\centering
\resizebox{0.45\textwidth}{!}{%
  \includegraphics{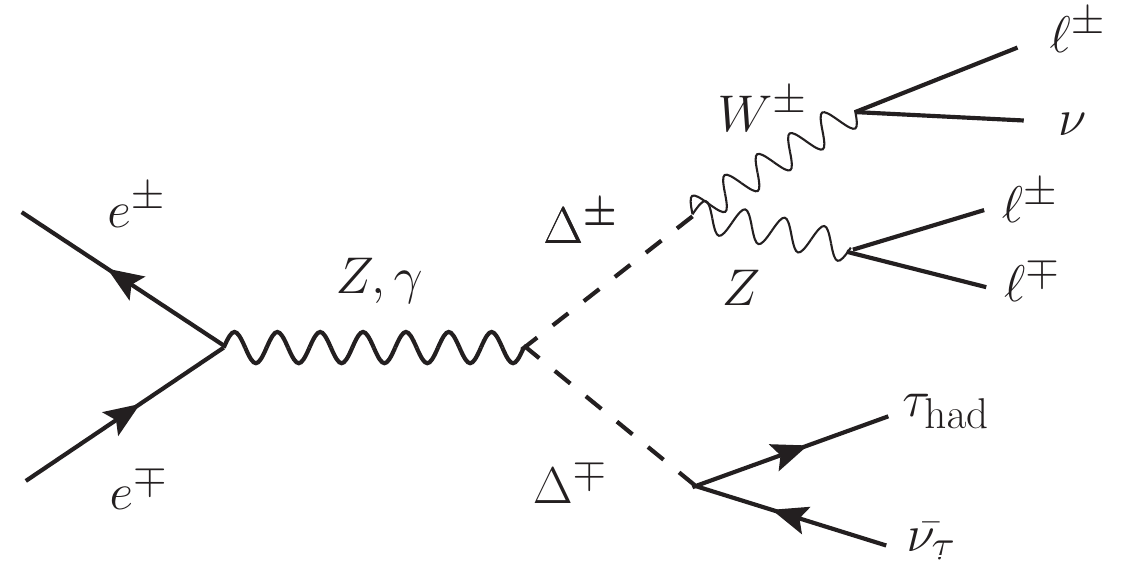}
}
\caption{Feynman diagram contributing to SR2.}
\label{fig:Feynmann_Sr2}
\end{figure}

This signal region is defined such that it is sensitive to  $e^+e^-\to (\Delta^\pm\to WZ\to 3\ell)(\Delta^\mp\to \tau\nu))$. We focus primarily on the hadronic decay of the $\tau$, mentioned as $\tau$-tagged jet ($\tau_\text{had}$), and the signal region is denoted as $\geq 3\ell + \tau_{\text{had}}$. Fig.~\ref{fig:Feynmann_Sr2} shows the corresponding Feynman diagram contributing to this SR. The basic requirement for selecting the  leptons and the hadronic $\tau_{\text{had}}$ are given in Table~\ref{tab:SRs}.

\begin{table}[t]
\centering
\resizebox{\columnwidth}{!}{%
\begin{tabular}{p{0.40\columnwidth} p{0.55\columnwidth}}
\toprule
Variables & Definition \\
\midrule
$m(\ell_1,\ell_2,\ell_3)$  & Invariant mass of leading 3-lepton system ($\ell_{123}$) \\
$p_\text{T}(\tau_j)/\sqrt{H_\text{T}}$ & Ratio of tau-tagged jet $p_\text{T}$ and the square-root of $H_\text{T}$ \\
$M_{\text{eff}}$ & Effective mass \\
$E(\ell_1)$ & Energy of leading lepton \\
$L_\text{T}$ & Scalar sum of all leptonic $p_\text{T}$ \\
$E(\ell_2)$ & Energy of sub-leading lepton \\
$E_{\text{eff}}$ & Effective energy ($= E(\text{jets}) + E(\text{leptons}) + $ $p^{\text{miss}}_\text{T}$) \\
$E(\ell_3)$ & Energy of $\ell_3$ \\
$E(\tau_j)$ & Energy of $\tau_j$ \\
$m(\ell_1,\ell_2)$ & Invariant mass of leading 2-lepton system ($\ell_{12}$) \\
\bottomrule
\end{tabular}%
}
\caption{The input variables used in the DNN training for signal region SR2.}
\label{tab:SR2_variables}
\end{table}

\begin{figure*}[htb!]
\centering
\resizebox{0.45\textwidth}{!}{%
  \includegraphics{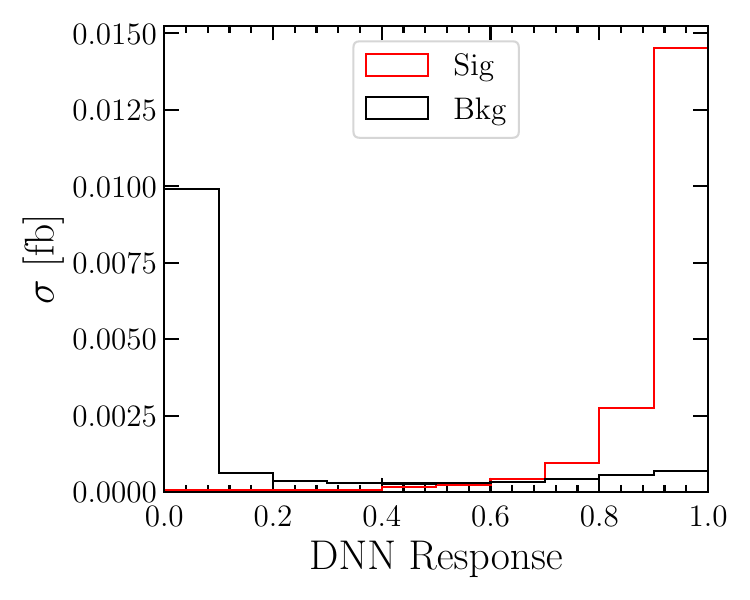}
}
\resizebox{0.45\textwidth}{!}{%
  \includegraphics{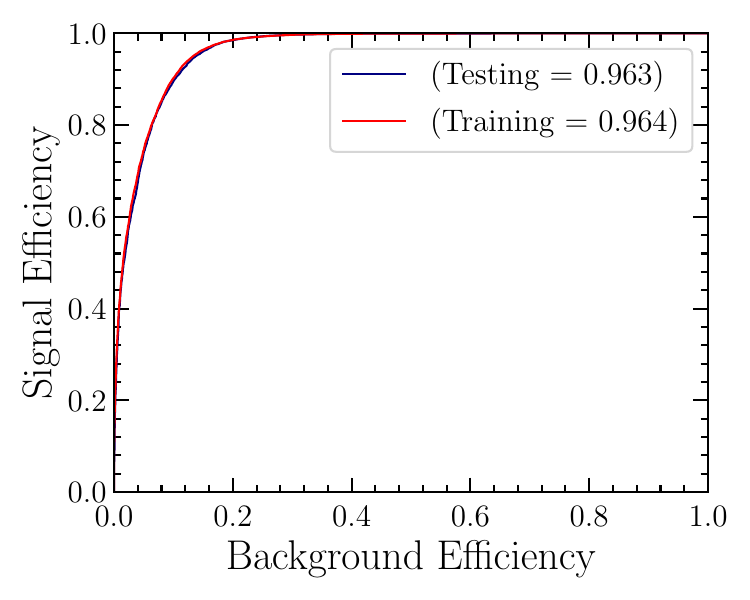}
}
\caption{Left: DNN response to signal and background for both training and testing samples, for SR2. Right: ROC-curve for both training and testing samples for SR2.}
\label{fig:SR2_DNNoutput}
\end{figure*}

\begin{table}[htb!]
\centering
\begin{tabular}{lcc}
\toprule
Process &  $\ge 3\ell + \tau_\text{had}$ & DNN Response $>$ 0.7\\
\midrule
Signal & 0.019 fb & 0.016 fb\\
VV $\tau \nu$ & 0.013 fb & 0.001 fb\\
\bottomrule
\end{tabular}
\caption{Fiducial cross sections  of both the signal and background after the preselction and DNN score for SR2, where $V = W^\pm, Z \text{ and } h$.}
\label{tab:SR2_cutflow}
\end{table}

We present the input features used for DNN for this signal region in Table~\ref{tab:SR2_variables}. We have considered a DNN with 3 hidden layers with 64, 32, and 16 neurons, respectively.  The DNN architecture is not the same as for SR1, since we optimise the DNN for this SR independently. Then, we divide the whole data set into a training and a testing dataset, with 80\% of the data in the training set and the rest in the testing set. The loss function is computed using the binary cross-entropy function, and the Adam optimiser is used to update weights by backpropagation. After the training, the model is tested with the rest of the data. The resulting DNN response and ROC curve are shown in Fig.~\ref{fig:SR2_DNNoutput}. The AUC scores are 96.3\% and 96.4\% for testing and training samples, respectively. From the DNN response, we considered a DNN score of 0.7 to be suitable for distinguishing signal from the background. The fiducial cross sections of both the signal and background after the above DNN score are given in Table~\ref{tab:SR2_cutflow}. For an integrated luminosity of $\mathcal{L} = 500 \text{ fb}^{-1}$ we get a discovery-level significance of $\approx$  $5\sigma$.

\subsection{SR3: $\ge 4j + \tau_{\text{had}}$}

\begin{figure}[htb!]
\centering
\resizebox{0.45\textwidth}{!}{%
  \includegraphics{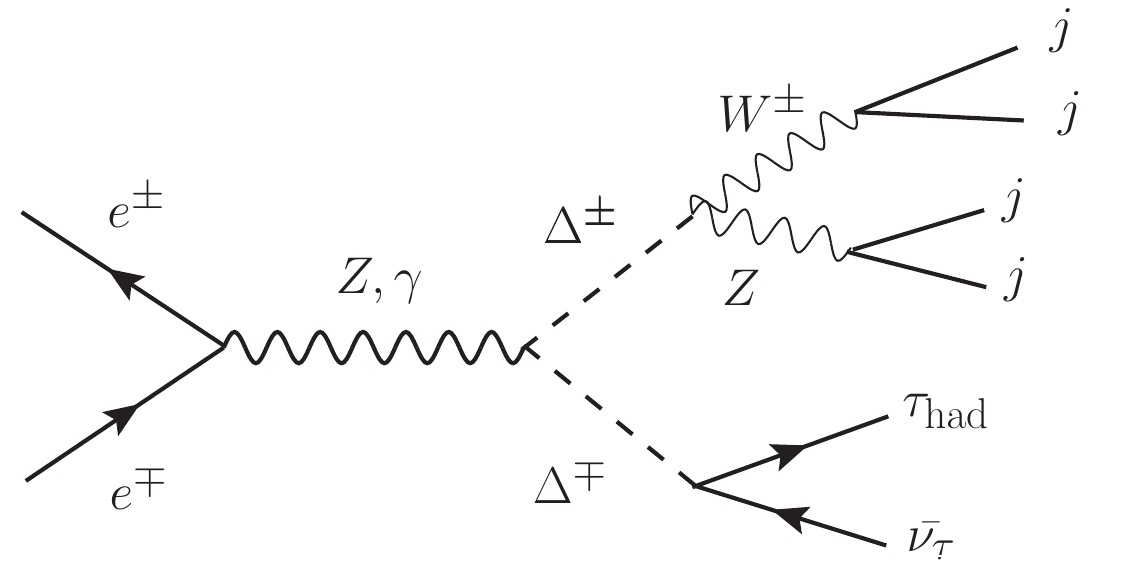}
}
\caption{Feynman diagram contributing dominantly to SR3.}
\label{fig:Feynmann_Sr3}
\end{figure}

This SR is defined such that the final states with hadronic $W^\pm Z$  and $\tau\nu$ decays contribute. Hence, we select a hadronic $\tau$($\tau_{\text{had}}$) along with at least 4 jets, and the signal region is denoted as $\geq 4j + \tau_{\text{had}}$. Since we forced one of the $\Delta^\pm$ to $\tau_{had} \nu$ mode, the one hadronic tau tagging helps selecting the required signal event. Again from the remaining jets we select pairs that reconstruct $W,Z$ with a 20 GeV mass window. Then with this reconstructed four vector, we combine any two jets from the remaining ones to reconstruct the charged scalar mass, However we select that combination which is closer to 150 GeV as our reconstructed charged scalar. Fig.~\ref{fig:Feynmann_Sr3} shows the dominant Feynman diagram for this SR.

\begin{table}[htb!]
\centering
\resizebox{\columnwidth}{!}{%
\begin{tabular}{p{0.40\columnwidth} p{0.55\columnwidth}}
\toprule
Variables &  Definition \\
\midrule
$p_\text{T}(j_1)$  &  Transverse momentum of $j_1$\\$p_\text{T}(j_2)$  &  Transverse momentum of $j_2$\\$p_\text{T}(j_3)$  &  Transverse momentum of $j_3$\\
$|\Delta\phi(W \text{ or }Z,\Delta^\pm)|$  &  Angular separation between $W^\pm$ or $Z$ and $\Delta^\pm$\\
$E_{\text{eff}}$  & Effective energy ($= E(\text{jets}) + E(\text{leptons}) + p^{\text{miss}}_\text{T}$)\\
\bottomrule
\end{tabular}%
}
\caption{Definition of the variables used in signal region SR3.}
\label{tab:SR3_variables}
\end{table}

\begin{table}[t]
\centering
\resizebox{\columnwidth}{!}{%
\begin{tabular}{lcc}
\toprule
Cuts &  Signal [fb] & Background [fb]\\
\midrule
After preselection & 1.3 & 1.25\\
$10~\text{GeV} < p_\text{T}(j_1) < 100~\text{GeV}$ & 1.28  & 1.17\\
$10~\text{GeV} < p_\text{T}(j_2) < 65~\text{GeV}$ & 1.272 & 1.07\\
$10~\text{GeV} < p_\text{T}(j_3) < 50~\text{GeV}$ & 1.27 & 1.03\\
$|\Delta\phi(W \text{ or } Z,\Delta^\pm)| > 2.0~\text{rad}$ & 1.253 & 0.92 \\
$E_{\text{eff}} < 240~\text{GeV}$ & 1.252 & 0.70\\
$m_{\Delta^\pm} < 200~\text{GeV}$ & 1.25 & 0.41\\
\bottomrule
\end{tabular}%
}
\caption{Cut-flow summary for SR3, showing the signal and background cross sections in femtobarn after each selection cut.}
\label{tab:SR3}
\end{table}

\begin{figure}[t]
\centering
\resizebox{0.45\textwidth}{!}{%
  \includegraphics{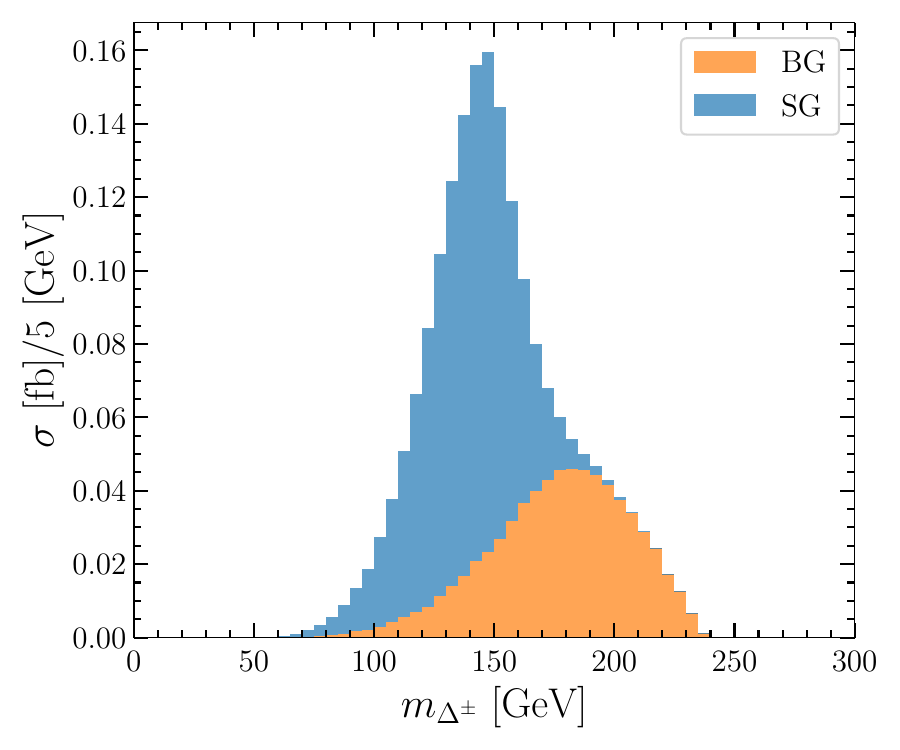}
}
\caption{Invariant mass reconstruction of $\Delta^\pm$ from SR3.}
\label{fig:invmassHpm}
\end{figure}

For this SR, we did not use a DNN since the signal distributions are quite distinct compared to the background, but performed a cut-based analysis. All relevant discriminating observables (see Table~\ref{tab:SR3_variables}) are shown in Fig.~\ref{fig:SR3Distribution} in Appendix~\ref{app:SR_Dists}. The fiducial cross sections after each cut are shown in Table~\ref{tab:SR3}. With the resulting values, we find that $5\sigma$ can be reached with an integrated luminosity $\mathcal{L}\approx$ 17\,$\text{fb}^{-1}$.

From Table~\ref{tab:SR3}, before the $m_{\Delta^\pm} < 200$ GeV cut, the visible cross section is 1.252 ${\rm fb}$, and for $\mathcal{L} = 500 {\rm fb}^{-1}$ we have 626 events. Performing a Gaussian fit to the invariant mass distribution, see Fig.~\ref{fig:invmassHpm}, the width of the distribution is 18.38 GeV. Hence the rough resolution of the charged scalar mass is $18.38/\sqrt{626} = 735$ MeV. Thus the charged scalar mass can be measured with an accuracy of less than 1\,GeV.\footnote{
The systematic uncertainty is dominated by the jet energy scale of $\approx 150$\,MeV~\cite{An:2018dwb,CEPCStudyGroup:2018ghi}.} To be more specific, the statistical uncertainty is dominant and equals $\approx 750$\,MeV for 500\,fb$^{-1}$ integrated luminosity.

\begin{figure}[H]
\centering
\resizebox{0.45\textwidth}{!}{%
  \includegraphics{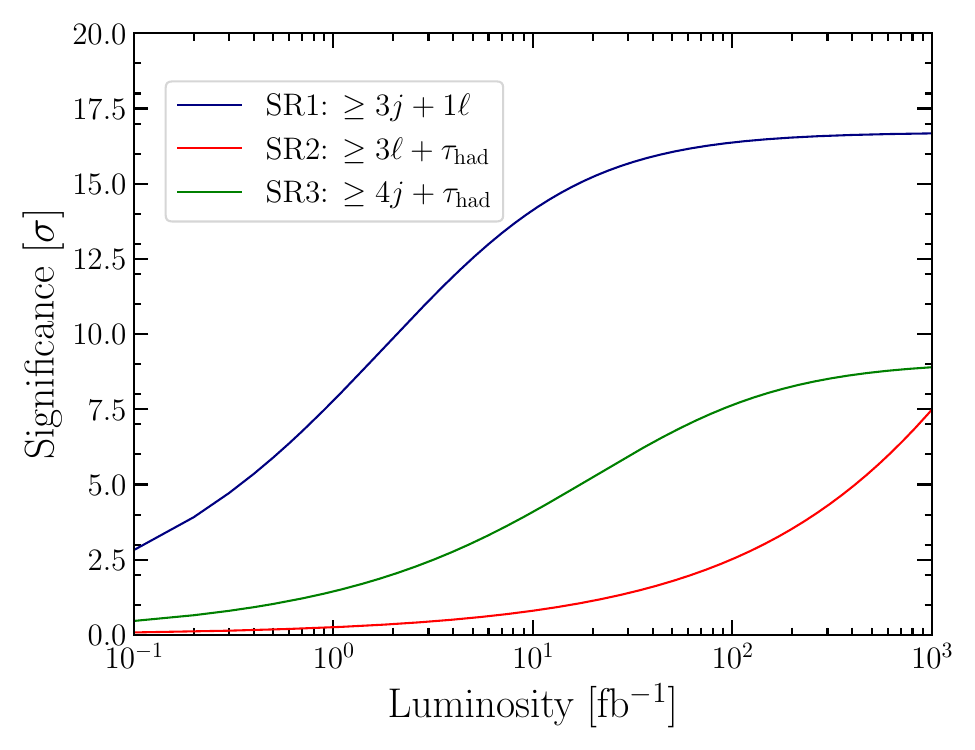}
}
\caption{Signal significance for $m_{\Delta^\pm} = 150$\,GeV as a function of integrated luminosity for the three defined signal regions: SR1, SR2, and SR3.}

\label{fig:final_result}
\end{figure}

Finally, Fig.~\ref{fig:final_result} shows the expected significances for the three signal regions. The leading channel is SR1, where significances above 10$\sigma$ can already be reached with a small dataset. The significance plateaus due to the impact of systematics. SR3 is sub-leading, while SR2 is statistics-limited and thus shows a different behaviour than the other two systematically limited signal regions. Final states with four or more electrons and muons can be studied further in future efforts. Also integrated luminosity $\mathcal{L} = 500 {\rm fb}^{-1}$, the significance that could be achieved for SR1 (SR2) [SR3] is $16.64\sigma\,(5.47\sigma)\,[8.72\sigma]$.

\section{Conclusions}\label{sec:conclusion}

In this article, we examined the discovery prospects for a light charged Higgs within the $Y=0$ Higgs triplet model, the $\Delta$SM, at future electron-positron colliders. Motivated by the multi-lepton anomalies, which predict a new neutral scalar with a mass of $150\pm5$\,GeV, and $\gamma\gamma, Z\gamma, WW$ excesses at  152$\pm$1\,GeV, we considered $m_{\Delta^\pm} = $ 150\,GeV as a benchmark point (since the $\Delta$SM predicts that the charged and the neutral component are quasi-degenerate in mass). This is important, since detecting the charged component $\Delta^\pm$ at the LHC is particularly challenging due to several factors: the decay mode $\Delta^\pm \to \tau^\pm \nu$ leads to final states which resemble supersymmetric stau signatures (excluding masses below 110\,GeV), while $WZ$ and $tb$ involve multiple soft leptons or low-energetic jets (only excluding the mass region between $\approx 200\,$GeV -- 220\,GeV ~\cite{Butterworth:2023rnw}). These limitations significantly reduce the LHC sensitivity to $\Delta^\pm$, thereby motivating the need for a cleaner experimental environment. 

At future $e^+e^-$ colliders, one will have a reduced background and improved control over kinematics, making them ideal for probing electroweak-scale scalars with soft final state leptons. Assuming a centre-of-mass energy of 350\,GeV, we constructed three representative signal regions based on the nature of the possible final states. The first region (SR1) is designed to yield a high signal cross section with inclusive selection criteria. The second region (SR2) targets final states with three or more leptons and a hadronic $\tau$, which are very clean. The third region (SR3) is specifically designed to enable full reconstruction of the charged scalar mass. 

We find that the SR1 provides the highest event rates such that a discovery is possible with an integrated luminosity of 0.3 $\text{fb}^{-1}$. SR3 allows for measuring the mass of $\Delta^\pm$, $m_\Delta^\pm = 150$ GeV, (as shown in Fig.~\ref{fig:invmassHpm}) with a statistical uncertainty of $\approx 750$\,MeV for an integrated luminosity of 500\,$\text{fb}^{-1}$. The sensitivity of all three regions is summarized in Fig.~\ref{fig:final_result}, where we show the statistical significance as a function of integrated luminosity. 

Our results demonstrate that future lepton colliders provide a promising avenue to discover the 150 GeV charged scalar $\Delta^\pm$ and, more broadly, to explore the scalar sector responsible for the observed multi-lepton anomalies. Once discovered, measurements of the branching ratios will be possible in the setting of the future $e^+e^-$ collider. The study of the sensitivity to branching ratio measurements is beyond the scope of this article, but constitutes a promising future avenue.

\section{Acknowledgments}

AC is supported by a professorship grant of the Swiss National Science Foundation (Grant No.\ PP00P21\_76884). SPM, SB, MK, RM and BM acknowledge the support of the Research Office of the University of the Witwatersrand and the support from the South African Department of Science and Innovation through the SA-CERN program, and the National Research Foundation.

\appendix

\section{ Functions for Decay Widths}
\label{app:LoopFunc}
The decay widths for the triplet-like charged Higgs boson $\Delta^\pm$ presented in the main text contain several kinematic functions from two-body and three-body phase space, as well as off-shell kinematics. For completeness, we collect below the explicit expressions of these functions, following the conventions of Refs.~\cite{Rizzo:1980gz,Keung:1984hn,Djouadi:2005gi,Djouadi:2005gj,Ashanujjaman:2024lnr}. These results are useful for numerical evaluation of the decay rates and for cross-checking analytical results. The definitions are given as follows:
\begin{align*}
&\beta(x,y) = (1-x-y)^2-4xy
\\
&\beta_{ff^\prime}(x,y) = \left[ (x+y)(1-x-y)-4xy \right] \times \sqrt{\beta(x,y)},
\\
&\beta_t(x,y) = \frac{y^2}{x^3}(4xy+3x-4y) \log \frac{y(x-1)}{x-y} \nonumber \\ &\hspace{1.5cm} +(3x^2-4x-3y^2+1) \log\frac{x-1}{x-y} -\frac{5}{2} \nonumber
\\
&\hspace{1.0cm} +\frac{1-y}{x^2} (3x^3-xy-2xy^2+4y^2) + y\left(4-\frac{3}{2}y\right),
\\
&G(x,y) = \frac{1}{12y}\Bigg[2\left(-1+x\right)^3-9\left(-1+x^2\right)y \nonumber \\ &\hspace{1.0cm}+6\left(-1+x\right)y^2-6\left(1+x-y\right)y\sqrt{-\beta(x,y)} \times \nonumber
\\
&\hspace{1.0cm} \Bigg\{\tan^{-1}\left(\frac{1-x+y}{\sqrt{-\beta(x,y)}}\right) + \tan^{-1}\left(\frac{1-x-y}{\sqrt{-\beta(x,y)}}\right)\Bigg\} \nonumber  \\ &\hspace{1.5cm} -3\left(1+\left(x-y\right)^2-2y\right)y\log x\Bigg],
\\
&H(x,y) = \frac{1}{4x \sqrt{-\beta(x,y)}} \nonumber \\ &\hspace{1.0cm} \left\{ \tan^{-1}\left(\frac{1-x+y}{\sqrt{-\beta(x,y)}}\right) +\tan^{-1}\left(\frac{1-x-y}{\sqrt{-\beta(x,y)}}\right) \right\} \nonumber \\ &\hspace{1.0cm} \times \Big\{-3x^3+(9y+7)x^2 -5(1-y)^2x+(1-y)^3\Big\}  \nonumber \\ &\hspace{0.8cm} +\frac{1}{24xy}\Big\{(-1+x)(2+2x^2+6y^2-4x-9y+39xy) \nonumber
\\
&\hspace{1.0cm} - 3y(1-3x^2+y^2-4x-2y+6xy)\log x\Big\}.
\end{align*}

\section{Normalized Distributions}
\label{app:SR_Dists}

\begin{figure*}[htb!]
\centering
\resizebox{0.8\textwidth}{!}{%
  \includegraphics{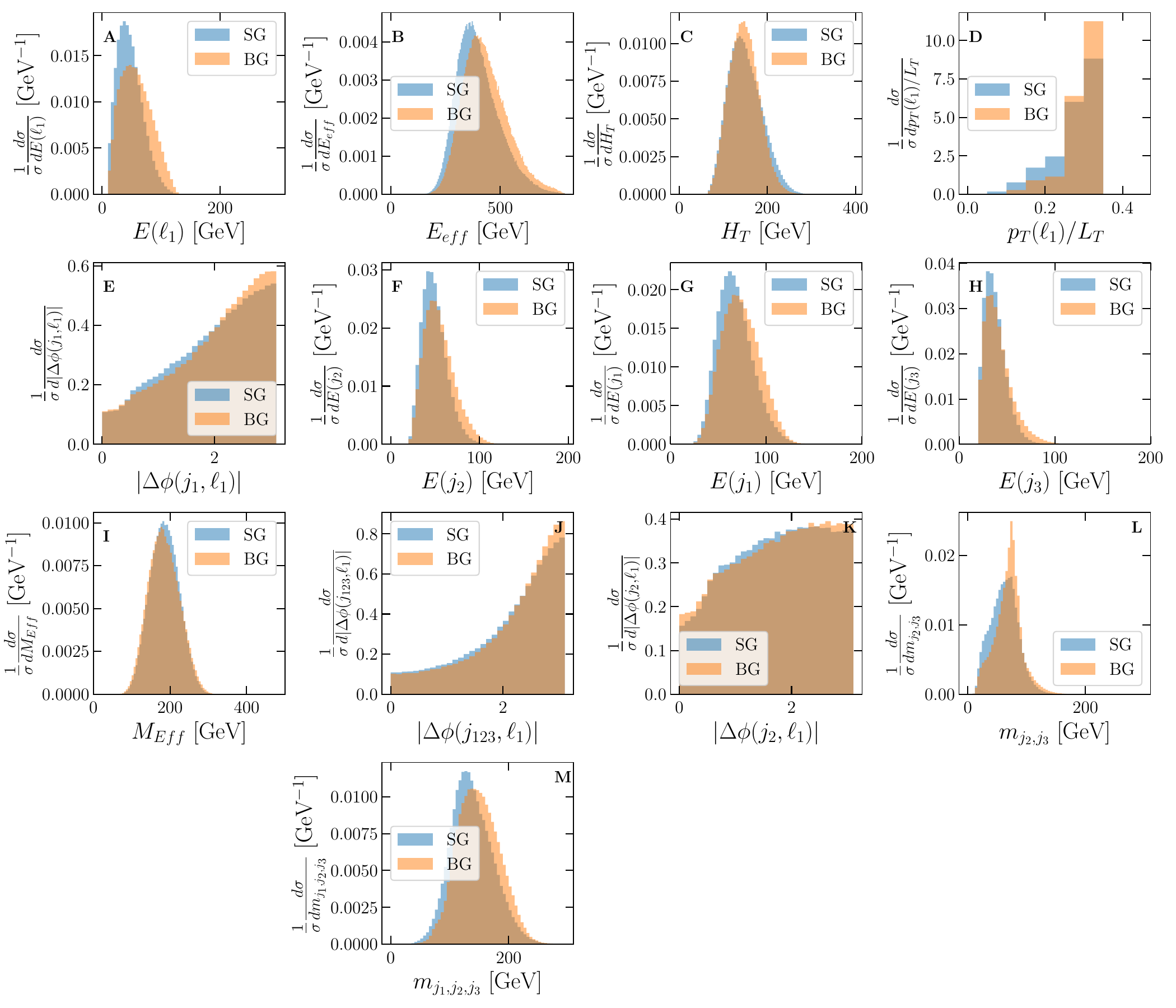}
}
\caption{Normalized distributions of key kinematic observables used in {{SR1}} to distinguish signal from background events.}
\label{fig:SR1Distribution}
\end{figure*}

\begin{figure*}[htb!]
\centering
\resizebox{0.8\textwidth}{!}{%
  \includegraphics{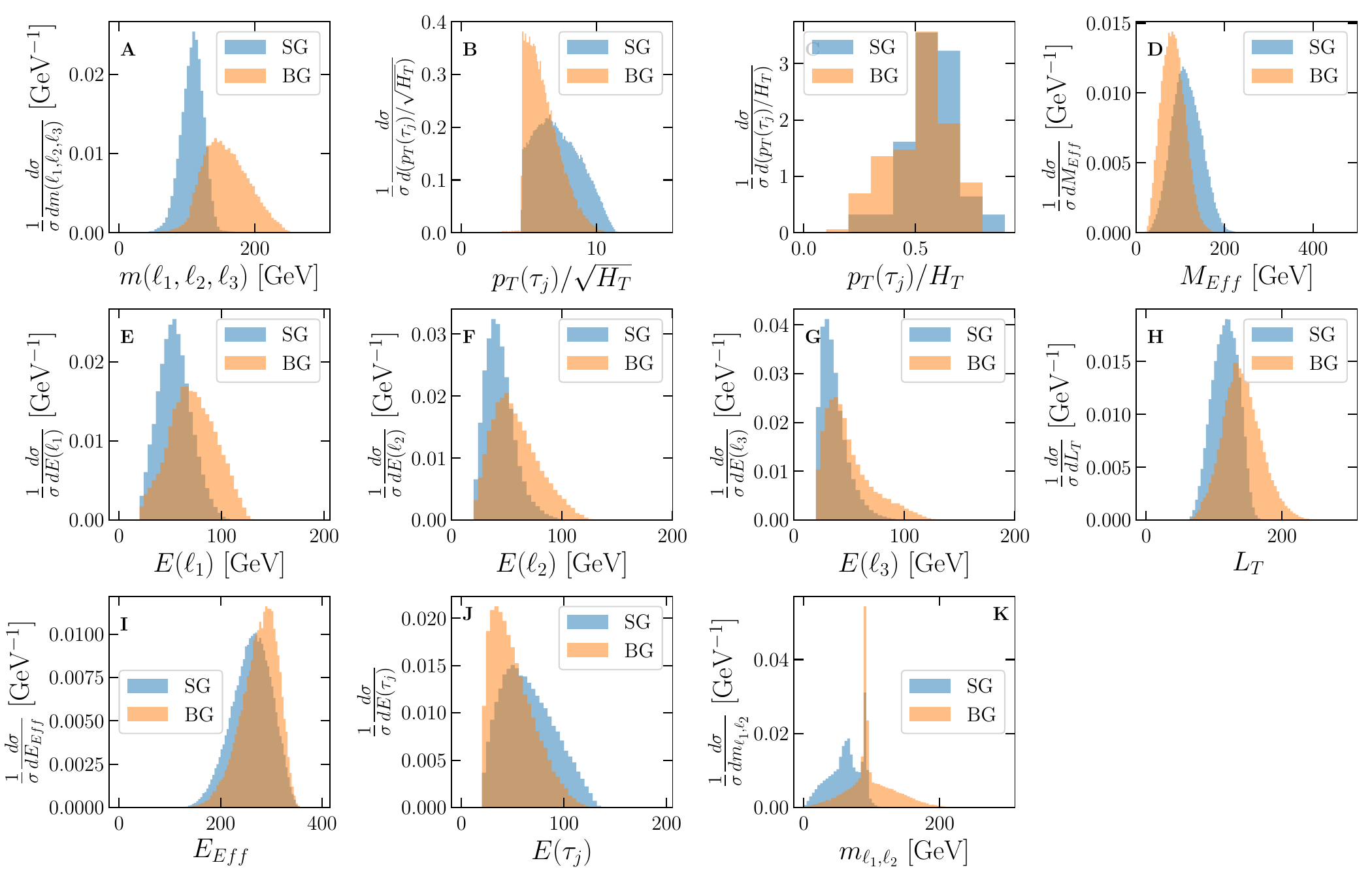}
}
\caption{Normalized distributions of key kinematic observables used in {{SR2}} to distinguish signal from background events.}
\label{fig:SR2Distribution}
\end{figure*}

\begin{figure*}[htb!]
\centering
\resizebox{0.8\textwidth}{!}{%
  \includegraphics{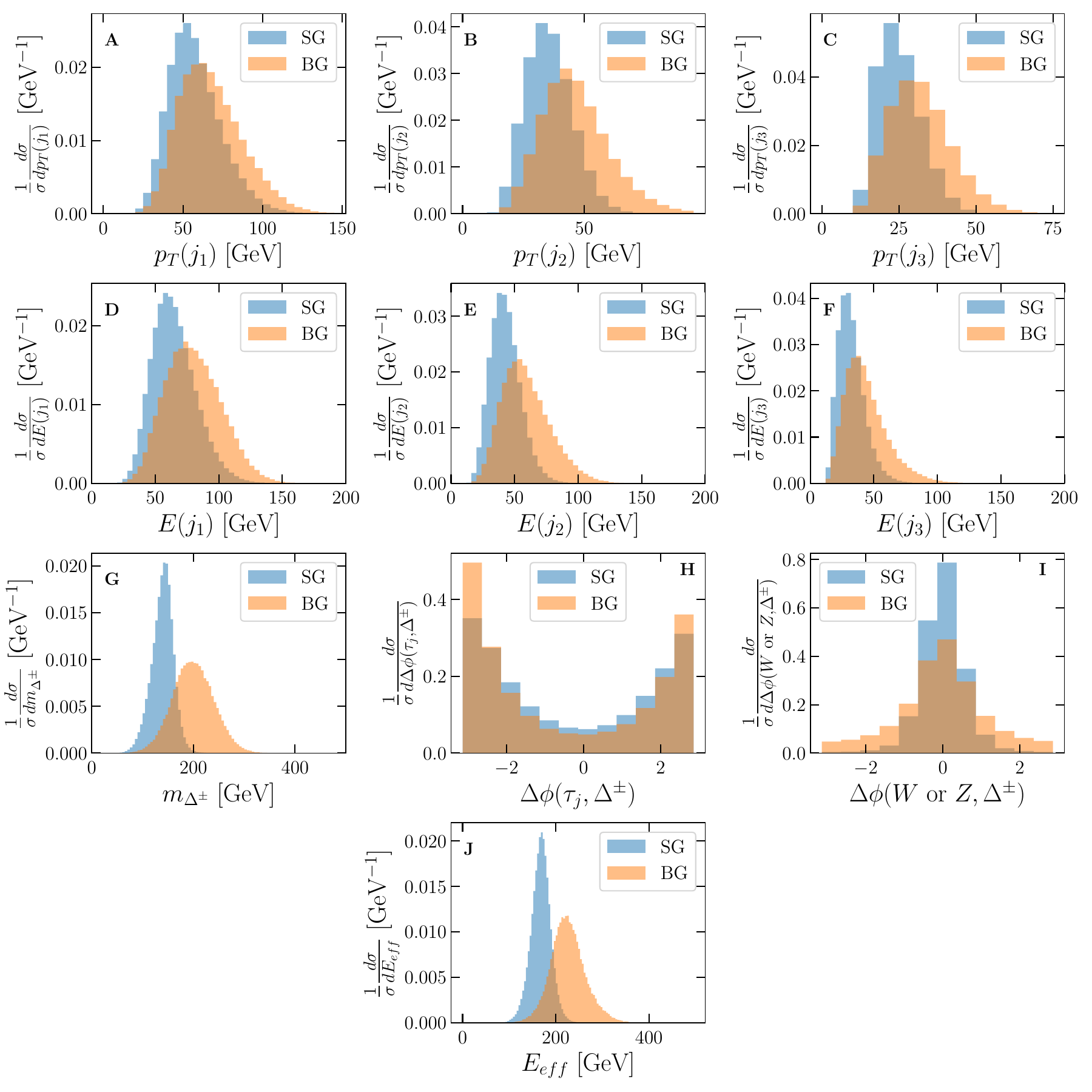}
}
\caption{Normalized distributions of key kinematic observables for {{SR3}} that are used to  distinguish signal from background events.}
\label{fig:SR3Distribution}
\end{figure*}

We provide the normalized distributions of key kinematic observables used in the analysis of the signal regions SR1, SR2, and SR3. These variables are crucial for distinguishing between the signal arising from $\Delta^\pm$ production and decay and the dominant SM backgrounds.
Fig.~\ref{fig:SR1Distribution}, Fig.~\ref{fig:SR2Distribution} and Fig.~\ref{fig:SR3Distribution} show the observables relevant for SR1, SR2 and SR3, respectively.

%
%
%
%
\bibliographystyle{utphys}
\bibliography{EPJC_format}
\end{document}